\documentclass[twocolumn,aps,amsmath,amssymb,floatfix,superscriptaddress,prb,longbibliography]{revtex4-1}
\usepackage{graphicx}
\usepackage{dcolumn}
\usepackage{bm}
\usepackage{xspace}
\usepackage{hyperref}
\usepackage{color}
\usepackage[flushleft]{threeparttable}

\def\ncto{Na$_2$Co$_2$TeO$_6$\xspace}
\def\nczto{Na$_2$Co$_{2-x}$Zn$_x$TeO$_6$\xspace}
\begin{document}

\title{Suppression of the antiferromagnetic order by Zn doping in a possible Kitaev material Na$_2$Co$_2$TeO$_6$}

\author{Zhongtuo~Fu}
\author{Ruokai~Xu}
\affiliation{Institute for Advanced Materials, Hubei Normal University, Huangshi 435002, China}
\author{Song~Bao}
\author{Yanyan~Shangguan}
\affiliation{National Laboratory of Solid State Microstructures and Department of Physics, Nanjing University, Nanjing 210093, China}
\author{Xin~Liu}
\author{Zijuan~Lu}
\affiliation{College of Physics and Electronic Science, Hubei Normal University, Huangshi 435002, China}
\author{Yingqi~Chen}
\author{Shuhan~Zheng}
\author{Yongjun~Zhang}
\author{Meifeng~Liu}
\author{Xiuzhang~Wang}
\author{Hong~Li}
\affiliation{Institute for Advanced Materials, Hubei Normal University, Huangshi 435002, China}
\author{Huiqian~Luo}
\affiliation{Beijing National Laboratory for Condensed Matter Physics, Institute of Physics, Chinese Academy of Sciences, Beijing 100190, China}
\affiliation{Songshan Lake Materials Laboratory, Dongguan, Guangdong 523808, China}
\author{Jun-Ming~Liu}
\affiliation{National Laboratory of Solid State Microstructures and Department of Physics, Nanjing University, Nanjing 210093, China}
\affiliation{Collaborative Innovation Center of Advanced Microstructures, Nanjing University, Nanjing 210093, China}
\author{Zhen~Ma}
\email{zma@hbnu.edu.cn}
\affiliation{Institute for Advanced Materials, Hubei Normal University, Huangshi 435002, China}
\affiliation{State Key Laboratory of Surface Physics and Department of Physics, Fudan University, Shanghai 200433, China}
\author{Jinsheng~Wen}
\email{jwen@nju.edu.cn}
\affiliation{National Laboratory of Solid State Microstructures and Department of Physics, Nanjing University, Nanjing 210093, China}
\affiliation{Collaborative Innovation Center of Advanced Microstructures, Nanjing University, Nanjing 210093, China}

\begin{abstract}
Very recently, a 3$d$ based honeycomb cobaltate \ncto has garnered tremendous attention due to the proposed proximity to the Kitaev spin-liquid state as its 4$d$/5$d$ counterparts. Here, we use Zn to substitute Co in a broad range and perform systematic studies on \nczto by structural, magnetic, and thermodynamic measurements, and track the doping evolution of its magnetic ground states. Due to the extremely close radii of Zn$^{2+}$ and high-spin Co$^{2+}$  ions, the substitution can be easily achieved. X-ray diffractions reveal no structural transition but only minor changes on the lattice parameter $c$ over a wide substitution range $0 \leq x \leq 1.5$. Magnetic susceptibility and specific heat measurements both suggest an antiferromagnetic ground state which is gradually suppressed with doping. It can survive with $x$ up to $\sim1.0$. Then it evolves into a spin-glass phase with short-range order that is rapidly supplanted by a magnetically disordered state when $x \geq 1.3$. By summarizing all these data, we construct a magnetic phase diagram of \nczto. Our results demonstrate that the Zn doping can effectively suppress the magnetic order and induce a possibe quantum paramagnetic state. These may serve as a platform to investigate the Kitaev physics in this system.
\end{abstract}

\maketitle

\section{Introduction}

Quantum spin liquids (QSLs) with fractional excitations and long-range quantum entanglement have been an attractive issue since the pioneering concept of resonating-valence-bond (RVB) model was proposed by Anderson in 1973~\cite{Anderson1973153}. It is believed that superconductivity in copper oxides emerges by doping a QSL~\cite{anderson1}, which will facilitate unveiling the mechanism of high-temperature superconductivity. RVB model was initially built upon the triangular lattice~\cite{Anderson1973153}, where spin exchange interactions cannot be satisfied simultaneously among different lattice sites due to the geometrical frustrations, leading to a macroscopically degenerate ground state with no static magnetic order~\cite{Anderson1973153,arms24_453,nature464_199,RevModPhys.89.025003,Broholmeaay0668}. However, strictly speaking, this ``model" is not a model and does not have an exact solution. In 2006, Kitaev constructed an exactly solvable QSL model with spin $S$ = 1/2 on the honeycomb lattice~\cite{aop321_2}, which is named the Kitaev QSL model. Unlike QSLs described by the RVB model, Kitaev model features bond-dependent anisotropic interactions (called Kitaev interactions) with an intrinsic frustration of the spin on the single site, which introduces strong quantum fluctuations and thus results in a magnetically disordered state~\cite{aop321_2,nrp1_264,TREBST20221}. More importantly, such a state can host fractional excitations represented by Majorana fermions that hold promise for the applications in quantum computation~\cite{Kitaev20032,aop321_2,RevModPhys.80.1083,Barkeshli722}.

The materialization of Kitaev model has been one of the central subjects in the field of strongly correlated electronic systems in recent years. It is guided by Jackli and Khaliullin's proposal in a Mott insulator with strong spin-orbit coupling (SOC)~\cite{prl102_017205} that stimulates extensive explorations in real materials with heavy $d^5$ transition metal ions~\cite{0953-8984-29-49-493002,nrp1_264,doi:10.1021/acs.chemrev.0c00641,TREBST20221}. Along this line, 5$d^5$ $A_2$IrO$_3$ ($A$ = Na, Li) family and 4$d^5$ $\alpha$-RuCl$_3$ are the representative candidates. Although they undergo antiferromagnetic transitions at low temperatures~\cite{npjqm4_12,doi:10.1021/acs.chemrev.0c00641}, which precludes them from being a Kitaev QSL, there is accumulating evidence that they are proximated to the Kitaev QSL state, and more importantly, there are dominant Kitaev interactions in these materials~\cite{nrp1_264,TREBST20221,PhysRevLett.118.107203,np11_462,PhysRevB.99.054426,PhysRevLett.108.127203,nm15_733,Banerjee1055,np13_1079}. Especially for $\alpha$-RuCl$_3$, it is found that the long-range zigzag magnetic order is fragile due to the competition between Kitaev and non-Kitaev interactions and can be suppressed by applying a moderate magnetic field within the honeycomb plane~\cite{PhysRevB.99.140413,PhysRevB.102.140402,nc10_2470,PhysRevLett.120.077203,PhysRevResearch.2.013014,nc12_4007,Xiaoxue_Zhao:57501}. However, there are still debates on the nature of the field-driven magnetically disordered phase~\cite{PhysRevLett.120.067202,PhysRevB.103.054440,PhysRevLett.120.117204,nature559_227,PhysRevLett.120.217205,np17_915,doi:10.1126/science.aay5551,np18_401,czajka2022planar,PhysRevLett.125.037202}.

More recently, the pursuit for Kitaev systems is extended to 3$d^7$ cobaltates with a high-spin electronic configuration of $t^5_{2g}e^2_g$. It has been proposed that Co$^{2+}$ ions under an octahedral crystal field of oxygens can give rise to pseudospin-1/2 degrees of freedom with Kitaev interactions~\cite{PhysRevB.97.014407,PhysRevB.97.014408,PhysRevLett.125.047201}, despite some skepticism that the weak SOC is insufficient to promote the compass interplay~\cite{PhysRevB.103.214447,PhysRevB.104.134425,https://doi.org/10.48550/arxiv.2204.09856}. Initially proposed candidates include honeycomb-layered magnets \ncto and Na$_3$Co$_2$SbO$_6$. Early experiments have revealed that the zigzag-type magnetic order in both of them~\cite{WONG201618,PhysRevB.94.214416,PhysRevB.95.094424}, as the cases of $d^5$ iridates and $\alpha$-RuCl$_3$~\cite{npjqm4_12,doi:10.1021/acs.chemrev.0c00641}. Further inelastic neutron scattering (INS) results show that these two cobaltates indeed host a spin-orbit assisted $J_{\rm eff}$ = 1/2 state characterized by the spin-orbit excitons observed at 20-28~meV~\cite{PhysRevB.102.224429,Kim_2021}, and low-energy spin dynamics can be described by a Heisenberg-Kitaev model even though it has not yet reached a consensus on the sign and magnitude of the Kitaev interactions~\cite{PhysRevB.102.224429,Kim_2021,PhysRevB.104.184415,Kim2021,nc12_5559,doi:10.1142/S0217979221300061}. Furthermore, a pure Kitaev QSL state might be achieved by applying an in-plane magnetic field in \ncto~\cite{PhysRevB.101.085120,PhysRevB.104.144426,nc12_5559}. All these results point to a promising direction for exploring Kitaev physics in this virgin ground.

In this work, we substitute magnetic Co$^{2+}$ with nonmagnetic Zn$^{2+}$ in \ncto in an extensive range and perform structural, magnetic, and thermodynamic studies to investigate the doping evolution of the magnetic ground states. Due to a comparable size between Co$^{2+}$ and Zn$^{2+}$ ions, the substitution can be easily achieved. X-ray diffractions~(XRD) reveal no structural transition but only minor changes on the lattice parameter $c$ within $0 \leq x \leq 1.5$. Magnetic susceptibility and specific heat measurements show a suppression of long-range magnetic order with increasing zinc content. After $x\sim1.0$, it develops into a spin-glass state with short-range order, which is rapidly supplanted by a magnetically disordered state when $x \geq 1.3$. These results explicitly track the evolution process of the magnetic ground states and  establish a magnetic phase diagram of \nczto. Zn doping may severe as a feasible way to enhance quantum fluctuations and induce quantum paramagnetic~(QPM) behaviors that may provide insights about the Kitaev physics.

\section{Experimental Details}

Polycrystalline samples of a series of compositions for \nczto with $0 \leq x \leq 1.5$ were synthesized by conventional solid-state reactions. The raw materials of dried Na$_2$CO$_3$ (99.99\%), Co$_3$O$_4$ (99.99\%), ZnO (99.99\%), and TeO$_2$ (99.99\%) powders with stoichiometric amounts were mixed and thoroughly ground. Then they were loaded into the alumina crucibles and sintered at 800~$^\circ$C in air for 108 hours. To obtain pure-phase compounds, intermediate grindings for several times were required. XRD data were collected at room temperature in an x-ray diffractometer (SmartLab SE, Rigaku) using the Cu-$K_\alpha$ edge with a wavelength of 1.54~\AA. In the measurements, the scan range of 2$\theta$ was from 10$^\circ$ to 90$^\circ$ with a step of 0.02$^\circ$ and a rate of 10$^\circ$/min. Rietveld refinements on XRD data were performed by the EXPGUI and GSAS programs. The dc magnetic susceptibility was measured within 2-300~K in a Quantum Design physical property measurement system~(PPMS, Dynacool), equipped with a vibrating sample magnetometer~(VSM) option. Specific heat was measured within 2-30~K in a PPMS Dynacool with applied magnetic fields up to 9~T.

\section{Results and discussions}

\begin{figure*}[htb]
\centerline{\includegraphics[width=6.8in]{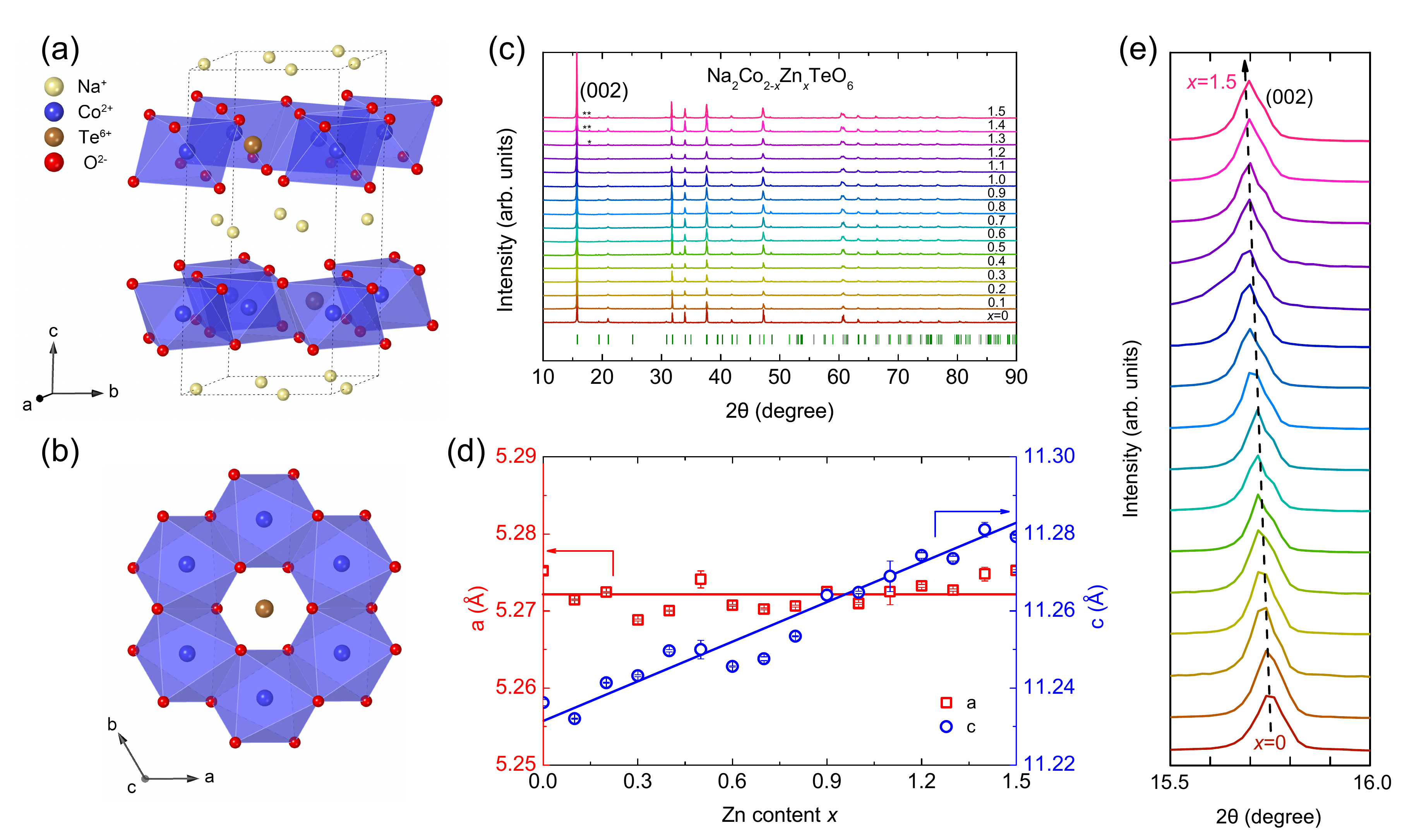}}
\caption{
(a) Schematic crystal structure of \ncto. (b) Top view of the honeycomb layer of CoO$_6$ octahedra. (c) X-ray diffraction~(XRD) patterns of \nczto with $0 \leq x \leq 1.5$. Green ticks denote Bragg peak positions of \ncto with space group $P6_322$ (No. 182). Asterisks mark the impurity peak of NaOH. (d) Evolution of the lattice parameters $a$ and $c$ with varying zinc concentration $x$. Solid lines are guides to the eye. (e) Zoom-in view of the most intense diffraction peak (002) for different $x$. The dashed line indicates a minor shift to the low-angle direction with increasing $x$. Errors represent one standard deviation throughout the paper.
\label{fig1}}
\end{figure*}

Figure~\ref{fig1}(a) shows the schematic of crystal structure of \ncto. It crystallizes into the hexagonal structure with space group $P6_322$ (No. 182)~\cite{VICIU20071060}. The magnetic honeycomb layers consist of Co$^{2+}$ and Te$^{6+}$ ions within the crystalline $a$-$b$ plane, which is illustrated more clearly in Fig.~\ref{fig1}(b). Six edge-shared CoO$_6$ octahedra form a regular honeycomb lattice with a Te$^{6+}$ cation located at the center. The honeycomb layers arrange along the $c$ axis and are well separated by an intermediate nonmagnetic layers of Na$^+$ ions. Since the nonmagnetic layers block Co-Co superexchange interaction pathways along the $c$ axis, the magnetic couplings are defined within the $a$-$b$ plane, featuring a quasi-two-dimensional magnetism~\cite{PhysRevB.94.214416,PhysRevB.95.094424}. In Fig.~\ref{fig1}(c), we show the XRD patterns of the Zn-substituted compound series \nczto as well as the Rietveld refinement results of the parent compound. For the $x$ = 0 compound, the refinements give lattice parameters $a = b$ = 5.2752(4)~\AA, $c$ = 11.2362(8)~\AA, and $\alpha = \beta =$ 90$^\circ$, $\gamma$ = 120$^\circ$, which are in line with existing literatures~\cite{VICIU20071060,PhysRevB.94.214416,PhysRevB.95.094424,doi:10.1021/acs.cgd.8b01770}. The refinement parameters are $R_{\rm p} \approx$ 2.68\%, $R_{\rm wp} \approx$ 3.58\%, and $\chi^2 \approx$ 1.598, respectively. With the zinc content $x$ increasing, there is no obvious peak splitting nor additional reflections except two very weak impurity peaks of NaOH between 17 and 18 degree appearing in high doping samples, excluding a structural transition and Zn residual in the regime even if $x$ is up to 1.5. Thus, we start with the original structure model of \ncto~\cite{doi:10.1021/acs.cgd.8b01770} and perform the refinements for compounds with $x>$~0 as well. It is found that almost all of the measured reflections can be well indexed with space group $P6_322$ and the impurity peaks in $x$ = 1.3 to 1.5 compounds have no significant influence on the sample quality. The nonmagnetic characteristic of NaOH further keeps the magnetic properties of this system free from the impact of impurity. The extracted lattice constants of $a$ and $c$ are depicted in Fig.~\ref{fig1}(d). One can see the in-plane Co-Co distance remains almost constant while the interlayer spacing slightly increases as $x$ gradually increases. In other words, the unit cell is a bit stretched along $c$ axis with Zn being introduced into the regime. Nevertheless, the overall change on the lattice parameter $c$ is less than 0.5\% over a wide substitution range $0 \leq x \leq 1.5$. Figure~\ref{fig1}(e) shows the zoom-in view of the most intense diffraction peak (002) for different $x$. There is indeed no splitting but a minor shift to low-angle direction with increasing $x$. The shift range is just within 0.05$^\circ$ despite $x$ increases up to 1.5. The continuous shift of this diffraction peak and monotonous increase of the lattice constant $c$ both imply the homogeneous substitution of Co with Zn in \nczto. In general, Co$^{2+}$ ions have two spin configurations that are low-spin and high-spin states, of which the ion radii are 0.65~\AA~and 0.745~\AA~for octahedral coordination~\cite{CHENAVAS19711057}, respectively. The former is a bit smaller compared with that of 0.74~\AA~of Zn$^{2+}$ while the latter is much closer to it~\cite{https://doi.org/10.1111/j.1151-2916.2003.tb03575.x}. Such a minor change with only 0.5\% on the lattice parameter $c$ upon so heavy zinc substitution is analogous to the cases of Zn$_{1-x}$Co$_x$TiO$_3$ and BaCo$_{1-x}$Zn$_x$SiO$_4$~\cite{https://doi.org/10.1111/j.1151-2916.2003.tb03575.x,2019Structural}, supporting the high-spin state of Co$^{2+}$ in \nczto.

\begin{figure*}[htb]
\centerline{\includegraphics[width=6.8in]{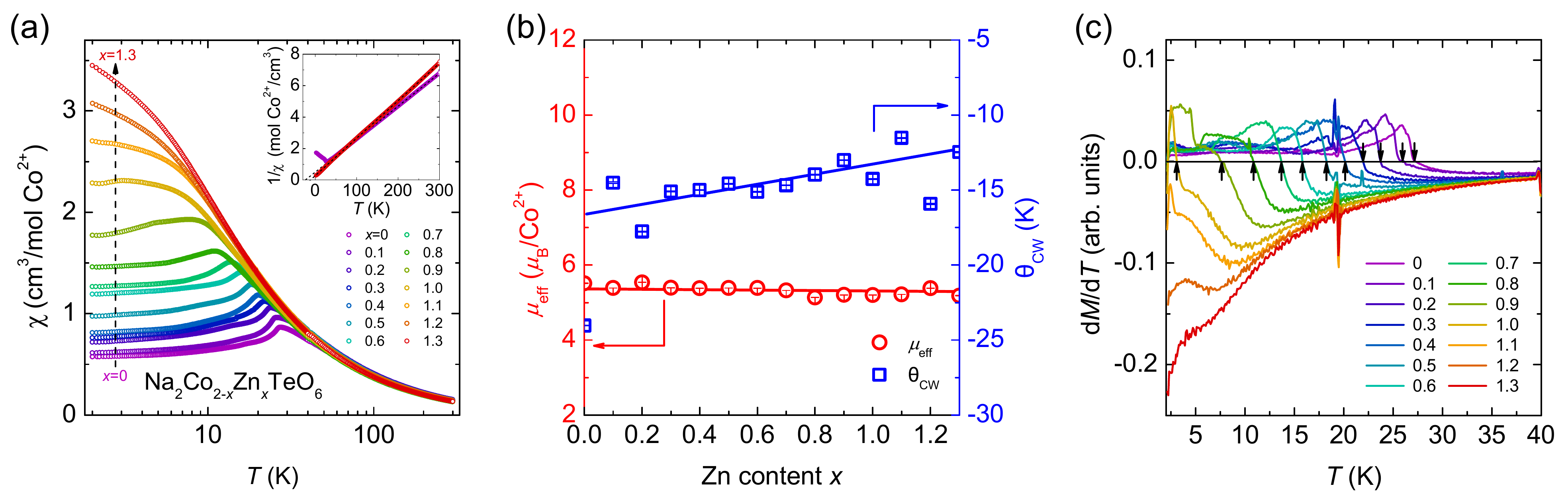}}
\caption{
 (a) Temperature dependence of the magnetic susceptibility~($\chi$) of \nczto with 0 $\leq x \leq$ 1.3. The inset shows the inverse susceptibility data of two representative zinc concentrations, which are $x$ = 0 and 1.3. Dashed lines are the fits with the Curie-Weiss law. (b) Extracted effective moment $\mu_{\rm eff}$ and Curie-Weiss temperature $\Theta_{\rm CW}$ as a function of $x$. Solid lines are guides to the eye. (c) Derivative of the magnetic susceptibility versus temperature for different $x$. Arrows mark the positions of d$M$/d$T$ = 0, from which $T_{\rm N}$ is extracted.
\label{fig2}}
\end{figure*}

We then characterize the compound series \nczto by measuring the dc magnetic susceptibility~($\chi$) with a field of 1~T, and the results are shown in Fig.~\ref{fig2}(a). For the $x$ = 0 compound, there is a sharp cusp at 27.1~K, suggestive of a phase transition from the paramagnetic~(PM) state towards an antiferromagnetic~(AFM) state upon cooling, consistent with other reports~\cite{PhysRevB.94.214416,PhysRevB.95.094424}. When introducing Zn$^{2+}$ into the system, the sharp cusp shifts to lower temperatures and becomes much broader until indiscernible at $x$$~\sim$~1.0, indicating an efficient suppression of the AFM phase transition by Zn substitution. We also notice an obvious rise in the magnetic susceptibility with increasing $x$ below the AFM transition temperature $T_{\rm N}$ = 27.1~K of the parent compound. This behavior is also observed in Ir-substituted $\alpha$-RuCl$_3$, which might result from the uncompensated moments triggered by nonmangetic impurities in the magnetically ordered state~\cite{PhysRevLett.119.237203}. When $T$ $>$ $T_{\rm N}$, the susceptibility data of various $x$ compounds are overlapped and they follow the Curie-Weiss behavior well [see the inset of Fig.~\ref{fig2}(a)]. From a fit of $\chi$ within 50 and 300~K to $\chi_{\rm 0}$ + $C/(T-\Theta_{\rm CW}$), where $\chi_{\rm 0}$, $C$, $\Theta_{\rm CW}$ denote temperature-independent term related to nuclear and Van-Vleck paramagnetic contributions, Curie-Weiss constant, and Curie-Weiss temperature, respectively, we obtain $\chi_{\rm 0} \approx$ -9.26~$\times$~10$^{-3}$~cm$^3$/mol Co$^{2+}$, $C \approx$ 3.80~cm$^3$~K/mol~Co$^{2+}$, and $\Theta_{\rm CW} \approx$ -24.03~K at $x$ = 0. Then it yields an effective moment $\mu_{\rm eff}$ = 5.52$~\mu_{\rm B}$/Co$^{2+}$, where $\mu_{\rm B}$ is the Bohr magneton. This value is much larger than that of 3.87~$\mu_{\rm B}$ for a spin-only $S$ = 3/2 effective moment, in favor of the high-spin state with unquenched orbital component in \ncto~\cite{PhysRevB.94.214416}. We also extract the values of $\mu_{\rm eff}$ and $\Theta_{\rm CW}$ for $x$ $>$ 0 compounds in this way, and the results are shown in Fig.~\ref{fig2}(b). The effective moment of Co$^{2+}$ remains almost constant even though $x$ increases up to 1.3. It means that Co$^{2+}$ maintains such a spin-orbit entangled state in the whole doping process. In the meanwhile, the decrease of $|\Theta_{\rm CW}|$ with increasing $x$ is observed, indicating a weakened strength of magnetic coupling. This should result from the block of Co-Co superexchange paths by the spin vacancies formed by Zn$^{2+}$ occupying Co$^{2+}$ sites. We also need to point out two details on the yield Curie-Weiss temperature. One is the drastic reduction of $|\Theta_{\rm CW}|$ upon doping, which should reflect an intrinsic behavior of this system, since all the physical measurements and data analysis of $x$~$>$~0 samples were performed under the same circumstances as the parent compound. The similar case is also observed in Na$_2$Ir$_{1-x}$Ti$_x$O$_3$ and it is attributed to the underlying quantum phase transition at quite a bit lower doping level~\cite{PhysRevB.89.241102}. We thus speculate that \nczto may undergo similar process. The other one is the very slow and even nearly constant change of $|\Theta_{\rm CW}|$ for the doped compounds. This insensitive response to nonmagnetic doping is also observed in Li$_2$Ir$_{1-x}$Ti$_x$O$_3$ and it is interpreted as an important signature of magnetic coupling beyond the nearest-neighbor interaction~\cite{PhysRevB.89.241102}. It should be true for \nczto that long-range spin interactions play a vital role in the spin Hamiltonian, even though the possible quantum phase transition as told above will significantly affect magnetic couplings of the parent compound, where a leading third-nearest-neighbor interaction in the Hamiltonian has been confirmed~\cite{PhysRevLett.129.147202}. The effective spin model upon doping remains to be determined by further investigations.

Since the sharp cusp in the $\chi-T$ curve at $x$ = 0 becomes much broader when zinc concentration approaches the moderate value, it is not easy to directly determine $T_{\rm N}$ based on the original data. To reveal the evolution of the magnetic ground state more accurately, we plot the differential susceptibility d$M$/d$T$ in Fig.~\ref{fig2}(c). The positions where d$M$/d$T$ = 0 are marked by the arrows and they actually correspond to the maxima in the original $\chi$-$T$ curves in Fig.~\ref{fig2}(a). One can see that the value shifts to low temperatures with increasing $x$ and completely disappears when $x \geq$ 1.1. This labels  $x$ = 1.1 as the ending concentration for the long-range AFM state, after which the system does not exhibit long-range order, and evolves into a spin glass with short-range order and magnetically disordered state. This behavior is also elucidated by our specific heat results that will be discussed next.

\begin{figure}[htb]
  \centering
  \includegraphics[width=0.98\linewidth]{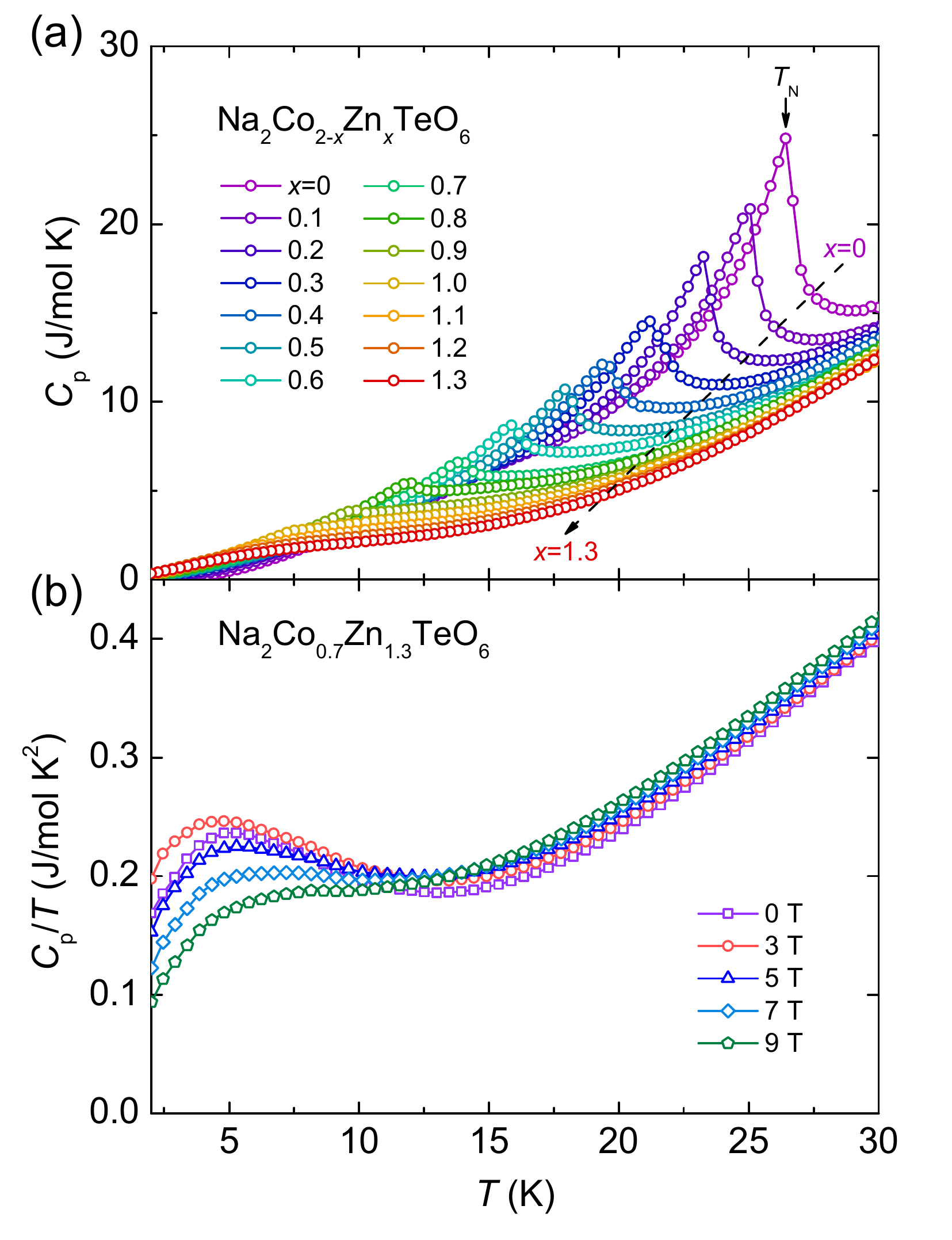}
  \caption{
  (a) Temperature dependence of the specific heat~($C_{\rm p}$) of \nczto within a temperature range of 2-30~K. The dashed line with an arrow denotes an evolution tendency of the antiferromagnetic~(AFM) phase transition with increasing $x$. (b) Temperature dependence of $C_{\rm p}/T$ of Na$_2$Co$_{0.7}$Zn$_{1.3}$TeO$_6$ in several different magnetic fields.}
  \label{fig3}
\end{figure}

To further verify the evolution of the magnetic ground states of \nczto, we perform specific heat measurements. In Fig.~\ref{fig3}(a), we present the results of specific heat at zero field for $0 \leq x \leq 1.3$. At $x$ = 0, there is a $\lambda$-shaped peak observed at 26.5~K, which is expected for an AFM phase transition~\cite{PhysRevB.94.214416,PhysRevB.101.085120}. The transition temperature is essentially in agreement with that of 27.1~K captured in the magnetic susceptibility. As $x$ increases, the sharp peak shifts to lower temperatures with intensity slightly weakened and then it evolves into a broad peak when $x$~$>$~1.0. Distinct from the sharp peak in $x~\leq$~1.0 compounds, the broad peak suggests that the system no longer hosts an AFM ground state with long-range magnetic order, but instead a magnetic state with short-range order or no any order~\cite{RevModPhys.58.801,RevModPhys.89.025003}. This result coincides with that obtained from the magnetic susceptibility. Subsequently, we select the $x$ = 1.3 compound as an example and perform measurements under several magnetic fields. The results of temperature dependence of $C_{\rm P}/T$ are illustrated in Fig.~\ref{fig3}(b). There is a more obvious broad peak captured at $\sim$5.1~K at zero field, which shifts to higher temperatures with increasing magnetic fields. Field dependence of the broad peak reflects an intrinsic behavior of the compound. This is an universal feature in some spin glasses and QSL candidates that may be associated with the development of short-range spin correlations~\cite{PhysRevB.98.220407,PhysRevLett.120.087201,prl98_107204}.

\begin{figure}[htb]
  \centering
  \includegraphics[width=0.98\linewidth]{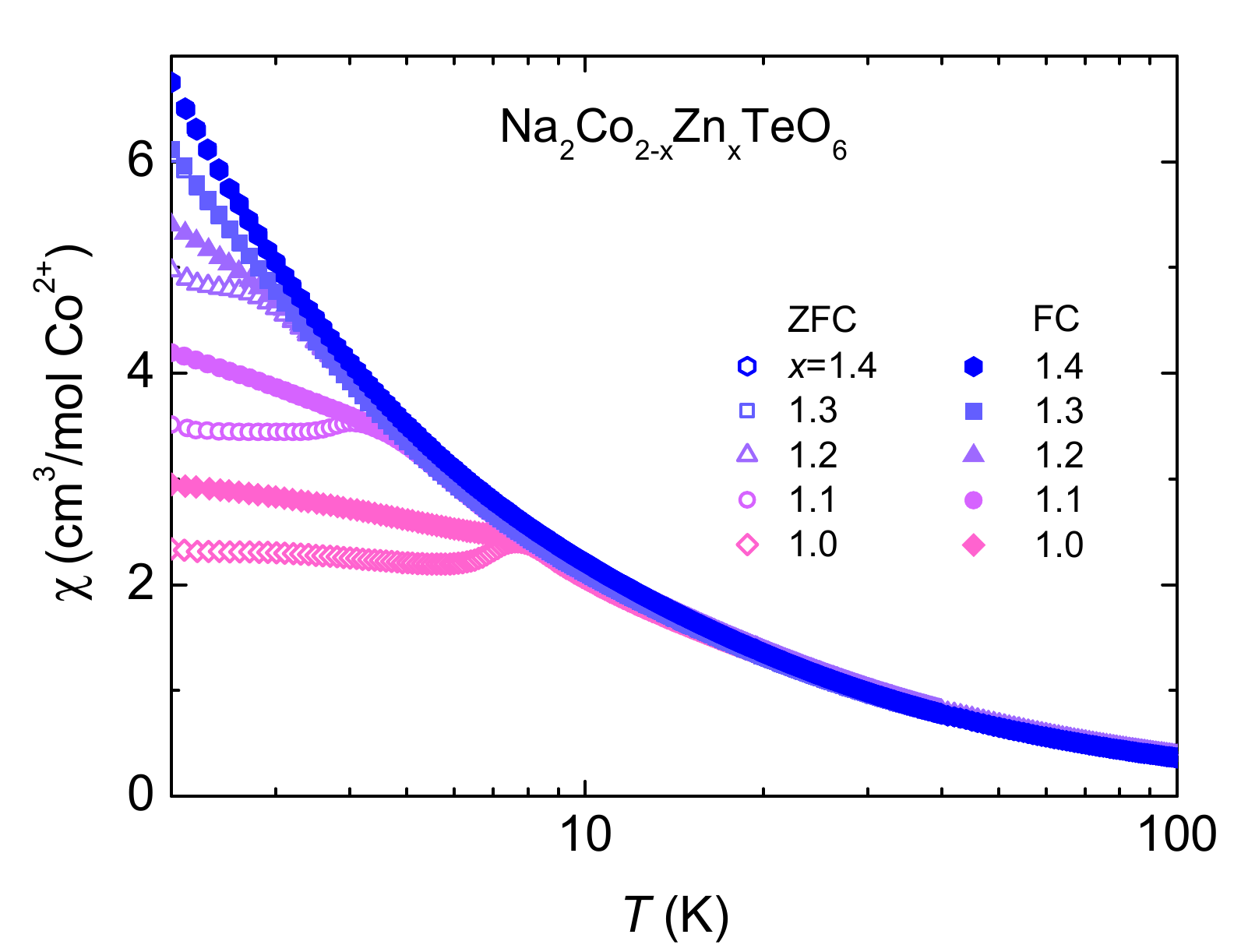}
  \caption{
  Temperature dependence of the magnetic susceptibility of \nczto when $1.0 \leq x \leq 1.4$. The data are collected under zero-field-cooling (ZFC, open symbols) and field-cooling (FC, filled symbols) conditions with a small magnetic field of 0.1~T.}
  \label{fig4}
\end{figure}

To elucidate the nature of the magnetic phase with no long-range order in heavy Zn-doped compounds, especially to distinguish spin glass and quantum disordered states, we measure magnetic susceptibility in both zero-filed-cooling (ZFC) and field-cooling (FC) conditions in a small field of 0.1~T. As shown in Fig.~\ref{fig4}, there is a bifurcation between ZFC and FC curves for $x$ = 1.0 to 1.2 compounds, which is a typical feature for a spin glass with frozen moments below the freezing temperature $T_f$~\cite{RevModPhys.58.801,Mydosh1993}. It is reasonable for a magnetically frustrated system with serious structural disorder resulting from the introduction of Zn$^{2+}$ ions, which is a main ingredient of a spin glass~\cite{RevModPhys.58.801,Mydosh1993}. Besides that, spin vacancies created by Zn$^{2+}$ substituting Co$^{2+}$ hinder the magnetic exchange paths and turn an antiferromagnet with long-range spin correlations to a short-range spin-glass phase when zinc content reaches a certain level. We also notice a slight upturn of the ZFC susceptibility data nearby the lowest temperature of 2~K. This Curie-like tail should be associated with some weakly interacting orphan spins generated via doping. Then the behavior of a bifurcation between ZFC and FC data rapidly disappears when $x \geq$ 1.3. In other words, the system evolves from the robust AFM state to an intermediate zone of spin-glass phase, and finally to a magnetically disordered state with Zn doping.

\begin{figure}[htb]
  \centering
  \includegraphics[width=0.98\linewidth]{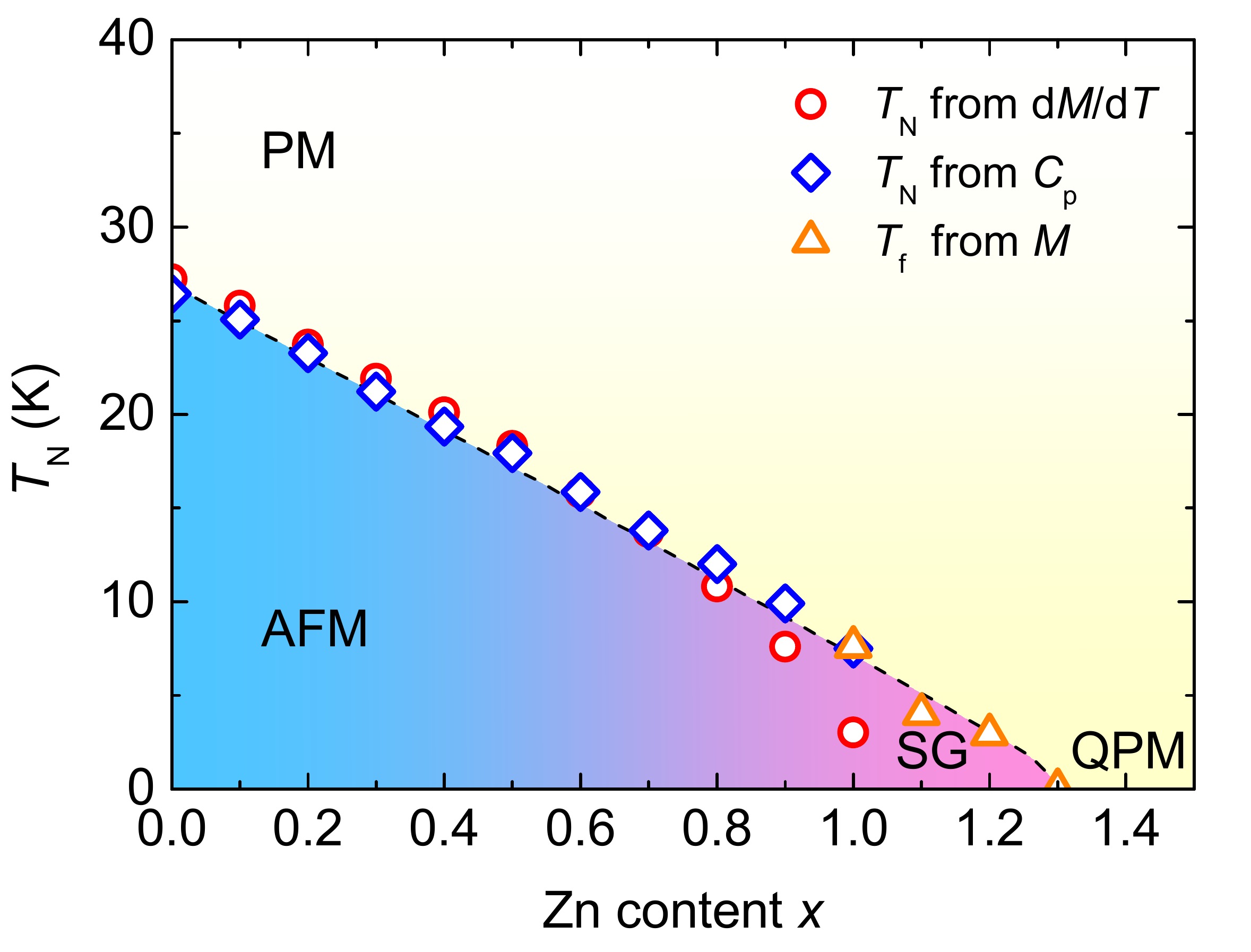}
  \caption{
  Magnetic phase diagram of \nczto. The high white zone represents the paramagnetic (PM) state. The bottom left blue zone and bottom right yellow zone denote AFM and quantum paramagnetic (QPM) states, respectively, where a spin-glass (SG) phase is located between them. The dashed line is a guide to the eye.}
  \label{fig5}
\end{figure}

After taking the aforementioned experimental data into account, we summarize the magnetic phase diagram of \nczto as illustrated in Fig.~\ref{fig5}. At $x$ = 0, the system undergoes a magnetic transition at $T_{\rm N}$~$\sim$~27~K with decreasing temperature~\cite{PhysRevB.94.214416,PhysRevB.95.094424}. When substituting Co$^{2+}$ with nonmagnetic Zn$^{2+}$, the transition temperature decreases continuously until vanishes at $x$~$\sim$~1.0. The phase boundary between AFM and PM is determined upon differential susceptibility d$M$/d$T$ = 0 as shown in Fig.~\ref{fig2}(c). $T_{\rm N}$ obtained in this way is consistent with the characteristic temperature where sharp $\lambda$-type peak occurs in the specific heat in Fig.~\ref{fig3}(a). As $x$ continues to increase, the long-range AFM state is completely destroy, followed by a spin-glass state with short-range order. Due to the coexistence of these two phases around an intermediate Zn doping, we cannot put forward an exact boundary to separate them. By contrast, the phase boundary between spin-glass and PM states can be pinned down by the freezing temperature $T_f$ between ZFC and FC susceptibility curves as shown in Fig.~\ref{fig4}. When $x \geq$ 1.3, there is no any anomalous behavior observed in either the magnetic susceptibility or the specific heat except for a broad peak in the latter, suggesting a magnetically disordered state. We speculate that this disordered state should be modulated by quantum fluctuations, since it shows the persistent spin dynamics down as low as 2~K. Eventually, the phase diagram of \nczto consists of four distinct phases, including a high-temperature PM state, and low-temperature AFM, spin-glass, and QPM states.

Previously, there have been some works reporting the doping effect in the widely studied Kitaev materials of 4$d^5$ $\alpha$-RuCl$_3$ and 5$d^5$ Na$_2$IrO$_3$. It is found that lightly Ir-doped $\alpha$-Ru$_{1-x}$Ir$_x$Cl$_3$ with $x$ = 0.2 and Ti-doped Na$_2$Ir$_{1-x}$Ti$_x$O$_3$ with $x$ = 0.05 can efficiently suppress zigzag-type AFM order and then drive them into a magnetically disordered state ($\alpha$-Ru$_{0.8}$Ir$_{0.2}$Cl$_3$) and a spin-glass state (Na$_2$Ir$_{0.95}$Ti$_{0.05}$O$_3$)~\cite{PhysRevLett.119.237203,PhysRevB.89.241102}. Compared with the 4$d$/5$d$ counterparts, the case in 3$d$ \nczto is different from both. AFM order in the parent compound is very robust with doping. It is completely destroyed until zinc content reaches a high level of $x$~$\sim$~1.0. The much higher percolation threshold where long-range AFM order is right absent may reflect a consequence of longer-distance magnetic coupling beyond the nearest-neighbor spin model in this 3$d$ cobaltate, such as the dominant third-nearest-neighbor magnetic interaction~\cite{PhysRevLett.129.147202}. Considering the case of Ru$_{1-x}$Ir$_x$Cl$_3$ where $ABC$-stacked zigzag order develops into a $AB$-stacked one at $x\sim$ 0.11, there should be an evolution of magnetic structures with Zn-doped content $x$ up to 1.0 in \nczto. The exchange interaction network will be also affected by different magnetic structures. It calls for further neutron diffraction experiment to elucidate this issue. In addition, there are three magnetic phases observed in \nczto at low temperatures, where the intermediate one of spin glass is sandwiched by the AFM and QPM states. It differs from the phase diagrams of Ru$_{1-x}$Ir$_x$Cl$_3$ and Na$_2$Ir$_{1-x}$Ti$_x$O$_3$ which only contain two parts at low temperatures~\cite{PhysRevLett.119.237203,PhysRevB.98.014407,PhysRevB.89.241102}. When the magnetic vacancies reaches a high level of $x\sim$ 1.0, they can effectively block the paths of long-range magnetic coupling and restrict it in a short-range zone like spin clusters. On the other hand, Zn content $x$ approaching to 1.0 actually corresponds to a half substitution of Co, where it has the most serious structural disorder effect. If the substitution is homogeneous, three among six Co sites in each honeycomb cell will be randomly occupied by Zn. In this case, the magnetic honeycomb lattice turns into a triangular lattice. It will further enhance the magnetically frustration effect in this system and thus the spin-glass phase with short-range spin correlations is formed. Although the QPM phase is revealed following the spin-glass state, is it a diluted QSL state as proposed in Ru$_{1-x}$Ir$_x$Cl$_3$~\cite{PhysRevLett.119.237203,PhysRevB.98.014407,PhysRevLett.124.047204}? It is hard to say with the current results. We need to point out that substitution will inevitably introduce site and bond disorder into the system, which may be detrimental to the realization of QSL states~\cite{PhysRevLett.119.157201,PhysRevLett.120.087201,PhysRevB.102.224415,Broholmeaay0668}, especially for the case of heavy doping in \nczto. Moreover, distinct from the sensitivity of AFM ground state to substitution in $\alpha$-Ru$_{1-x}$Ir$_x$Cl$_3$, such a high zinc content in \nczto can drive the regime approximate to be a nonmagnetic one that is more likely to exhibit trivial magnetic behaviors. Despite all these, the scenario of a heavily diluted QSL state is recently proposed in a triangular-lattice system NaYb$_{1-x}$Lu$_x$S$_2$~\cite{PhysRevMaterials.6.046201}. In the meanwhile, there are also theoretical proposals pointing out that diluted Kitaev materials may serve as potential candidates to host a variety of QSL phases~\cite{PhysRevLett.122.167202,PhysRevLett.124.047204,PhysRevX.11.011034,PhysRevLett.129.037204}. Thus, we cannot rule out the possibility of a QSL state in $x \geq$ 1.3 compounds. It is an open issue and remains to be investigated by more experimental and theoretical works in the future.

\section{Summary}

To summarize, we have performed XRD, magnetic susceptibility, and specific heat measurements on the Zn-substituted honeycomb coblatate \ncto that is proposed to be a Kitaev magnet. Due to the quite close radii of Zn$^{2+}$ and high-spin Co$^{2+}$ ions, there are minor changes on the lattice parameters even if the substitution content $x$ is up to 1.5. Magnetic susceptibility and specific heat results show a suppression of the AFM order as well as the magnetic coupling strength with $x$ increasing to 1.0. Then the system enters a spin-glass state with short-range order that is rapidly supplanted by a QPM state when $x \geq$ 1.3. These results show the evolution of the magnetic ground states with Zn doping, based on which we sketch a magnetic phase diagram of \nczto. Whether the QPM state upon heavy doping is related to QSL state deserves further explorations.

\section{Acknowledgements}

The work was supported by the National Key Projects for Research and Development of China with Grant No.~2021YFA1400400, National Natural Science Foundation of China with Grants No.~12204160, No.~12225407, No.~12074174, No.~12074111, No.~12204159, No.~12061130200, No.~11961160699, No.~11974392, and No.~12274444, Hubei Provincial Natural Science Foundation of China with Grant No.~2021CFB238 and No.~2021CFB220, China Postdoctoral Science Foundation with Grants No.~2022M711569 and No.~2022T150315, Jiangsu Province Excellent Postdoctoral Program with Grant No.~20220ZB5, K. C. Wong Education Foundation with Grant No.~GJTD-2020-01, the Youth Innovation Promotion Association of the CAS with Grant No.~Y202001, Beijing Natural Science Foundation with Grant No.~JQ19002, and Fundamental Research Funds for the Central Universities. Z.M. thanks Beijing National Laboratory for Condensed Matter Physics for funding support.

%

\end{document}